\documentclass[sigconf]{acmart}

\AtBeginDocument{%
  \providecommand\BibTeX{{%
    \normalfont B\kern-0.5em{\scshape i\kern-0.25em b}\kern-0.8em\TeX}}}

\acmConference[SE 2030]{International Workshop on Software Engineering in 2030}{November 2024}{Porto de Galinhas (Brazil)}
\acmISBN{978-1-4503-XXXX-X/18/06}

\usepackage{cleveref}

\begin{document}

\title{Innovating for Tomorrow: The Convergence of SE and Green AI}

\author{Luís Cruz}
\email{L.Cruz@tudelft.nl}
\orcid{0000-0002-1615-355X}
\affiliation{%
  \institution{Delft University of Technology}
  \country{The Netherlands}
}

\author{Xavier Franch Gutierrez}
\email{xavier.franch@upc.edu}
\orcid{0000-0001-9733-8830}
\affiliation{%
  \institution{Universitat Politècnica de Catalunya}
  \city{Barcelona}
  \country{Spain}}

\author{Silverio Martínez-Fernández}
\email{silverio.martinez@upc.edu}
\orcid{0000-0001-9928-133X}
\affiliation{%
 \institution{Universitat Politècnica de Catalunya}
 \city{Barcelona}
 \country{Spain}
}

\renewcommand{\shortauthors}{Cruz et al.}

\begin{abstract}
  The latest advancements in machine learning, specifically in foundation models, are revolutionizing the frontiers of existing software engineering (SE) processes. This is a bi-directional phenomona, where 1) software systems are now challenged to provide AI-enabled features to their users, and 2) AI is used to automate tasks within the software development lifecycle.
  In an era where sustainability is a pressing societal concern, our community needs to adopt a long-term plan enabling a conscious transformation that aligns with environmental sustainability values. In this paper, we reflect on the impact of adopting environmentally friendly practices to create AI-enabled software systems and make considerations on the environmental impact of using foundation models for software development.  
\end{abstract}



\maketitle

\section{Introduction}


Software-related CO\textsubscript{2} emissions from the ICT sector currently account for 2.1\%–3.9\% of global emissions~\cite{freitag2021real}. In today's context with the widespread use of AI systems, there have been many calls from industry leaders and AI experts admitting a further increase of these emissions. OpenAI chief executive Sam Altman warned that the next wave of generative AI systems will consume vastly more power than expected~\cite{crawford24}.
Hugging Face Climate Lead Sasha Luccioni has shown the scalability impact of AI systems' inference: ``While inference on a single example requires much less computation than that required to train the
same model, inference happens far more frequently than model training — as many as billions of times a day for a model powering a popular user-facing product such as Google Translate''~\cite{luccioni2023power}. Focusing only on AI systems, in a middle-ground scenario, by 2027 AI servers could use between 85 to 134 terawatt hours (Twh) annually~\cite{de2023growing}. That is similar to the annual electricity usage of countries such as Argentina, the Netherlands and Sweden individually and constitutes approximately 0.5 \% of the world's current electricity consumption~\cite{nyt23}.



The convergence of Software Engineering (SE) and AI is a bi-directional phenomena raising key sustainability concerns. On the one hand, software systems of diverse domains are now challenged to provide AI-enabled features to their users. To reduce the emissions from these AI systems from their development and training, to their usage and inference, and to its retirement, emerging AI software development lifecycles shall incorporate energy-aware capabilities. On the other hand, the potential of using AI and foundation models to automate tasks within the software development lifecycle is very promising, with dedicated events such as AIware\footnote{\url{https://2024.aiwareconf.org/} (visited on April 5, 2024)} and LLM4Code \footnote{\url{https://llm4code.github.io/} (visited on April 5, 2024)}. To fully embrace these technologies, the concerns regarding the emitted emissions from its usage shall be understood.


Therefore, the objective of this position paper is twofold:
\begin{enumerate}
    \item Identifying the trans-disciplinary dimensions and dichotomies in which the research from the SE community shall contribute to build greener AI systems, as well as reasoning on the evolution of SE practices in such dimensions and dichotomies. 
    \item Discussing the environmental sustainability of one application domain of AI systems: generative AI for SE tasks like generation of requirements, architecture, or code in which humans and intelligent agents jointly create software.
\end{enumerate}


\section{Overview of Green AI}

We define \textbf{Green AI} as a trans-disciplinary field that aims to make AI systems environmentally sustainable. Environmental sustainability of software (including AI systems) refers to engineering systems having minimal impact in our planet throughout its whole lifecycle \cite{becker2015sustainability}. We distinguish this from AI for Sustainability or AI for Green, where AI is used to make different domains more environmentally friendly – e.g., using AI to make agriculture more sustainable. We argue that it is important to make a clear separation between Green AI and AI for Green as they require different approaches and different scientific backgrounds. 

\begin{figure}
    \centering
    \includegraphics[width=0.9\linewidth]{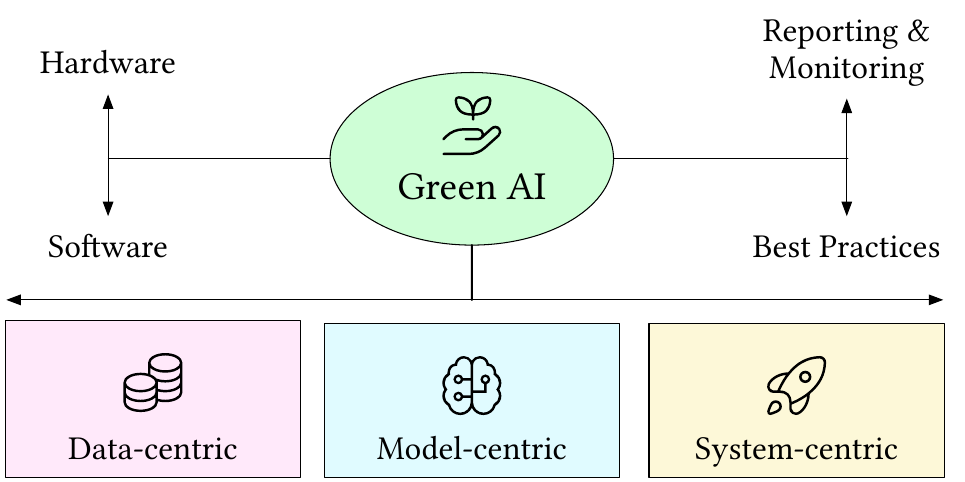}
    \caption{Overview of Green AI}
    \label{fig:green-ai}
\end{figure}

Given the complexity of AI systems, Green AI needs to be tackled from different angles, having each of these angles an important contribution to the overall carbon emissions of the systems.
Data, AI models, and the code of the software systems are the foundations of AI systems \cite{cd4ml19}. As such, to build and maintain sustainable AI systems, we shall focus in each of these three aspects. As depicted in Figure~\ref{fig:green-ai}, we divide Green AI across three major dimensions: data-centric, model-centric, and system-centric.
These dimensions are pinpointed below in \Cref{sec:data,sec:model,sec:system}. 
The figure also intersects Green AI with two dichotomies: hardware vs. software and reporting \& monitoring vs. best practices. 

\paragraph{Hardware $\leftrightarrow$ Software}

With the current trend of specializing hardware for particular AI tasks, practitioners can no longer be agnostic of the hardware they use to run their software. The choice of hardware is essential to ensure that a particular AI pipeline runs properly with no issues downstream. Hence, practitioners now face the challenge of having to be hardware experts. Moreover, previous research demonstrates how different hardware configurations can significantly impact the energy consumption of training deep-learning models~\cite{del2023dl}. However, these differences are not obvious and require meticulous trade-off analysis, which is sub-optimal in real-world scenarios.
Hardware design for AI poses more concerns. Since AI chips are specialized to particular AI tasks, updates to AI models might lead to different hardware requirements. This can significantly reduce the lifetime of hardware, posing challenges to the emerging ``Right to repair'' legislation efforts being discussed by the EU, the US, the United Nations and so on. Not surprising, the waste created by disposing hardware devices -- coined as e-waste -- is already a pressing environmental problem, as reported by the technology activist Murzyn\footnote{Murzyn is on a mission to report the hazards from handling electronic waste, a responsibility that has been delegated to Global South countries: \url{https://murzyn.asia/en/mission-00-24-42/} (visited on April 5, 2024)}. 


\paragraph{Reporting \& Monitoring $\leftrightarrow$ Best Practices} It is important that software systems are designed in a way that it is possible to report and monitor sustainability indicators~\cite{klas2009cqml}).
Practitioners need to provide accurate information about the carbon footprint of their experiments and models. This can be done e.g. via meta-data in repositories such as Hugging Face or using more elaborated and structured templates such as the concept of ID-card~\cite{SallamTOSEM}.
On the other side of the spectrum, we have best practices~\cite{jarvenpaa2023synthesis}. They provide practitioners with strategies that can be adopted to reduce the environmental footprint of their software.

These two poles complement each other. Effective best practices can only be established when reliable measurement mechanisms are in place. Monitoring and reporting not only incentivize practitioners to adopt these best practices but also provide essential feedback on their implementation and effectiveness.

\subsection{Data-centric Green AI}\label{sec:data}

Data-centric Green AI revolves around preparing data in a way that is expected to reduce the overall energy footprint of AI systems without hindering their performance.
Previous work has addressed this from multiple ways. Reducing data dimensionality, using feature selection or stratified random sampling already leads to linear gains in energy consumption~\cite{verdecchia2022data}.

Further strategies are being explored. 
Active learning \& coreset extraction have promising potential by identifying the most valuable examples that contribute to the learning process and excluding redundant or non-informative ones with confidence~\cite{salehi2023data}.

Knowledge transfer/sharing consists of using knowledge rather than solely relying on data to train a model. For instance, existing pre-trained models can be used to train new models. Preliminary research showcases improvements in energy consumption by a factor of 15 with this technique~\cite{walsh2021sustainable}.

Dataset distillation or dataset condensation involves synthesizing a smaller dataset derived from the original dataset, aiming to train a model on this reduced set while achieving test accuracy comparable to that of a model trained on the original dataset. This approach differs from coreset extraction by focusing on synthesizing informative samples rather than selecting existing ones. Data distillation has been used to reduce the size of the MNIST image set~\cite{lecun2010mnist} into just 10 synthetic distilled images (one per class) and achieve close to original performance~\cite{wang2018dataset}.

Curriculum learning consists of presenting training examples to the model in a specific order, starting
with simpler examples and gradually increasing the difficulty of the examples as training progresses. This strategy provides the model with a structured learning experience that mimics how humans learn and requires less iterations to converge, with time reductions of 70\%~\cite{platanios2019competence}.

\subsection{Model-centric Green AI}\label{sec:model}
Model-centric Green AI means developing experimental research to improve the AI model performance and energy efficiency. The target is to build and optimize AI models that can achieve similar outcomes while requiring fewer resources.

A notable example is the BLOOM model, whose carbon footprint have been studied during the training and inference stages to compete with other Large Language Models (LLMs) \cite{luccioni2023estimating}.
A more recent example are small language models (SLMs), like ``Phi'' that has shown performance comparable to models 5x larger \cite{slm-microsoft23}. Indeed, there is evidence that there is not always a trade-off between green and performance metrics. A mining repository study on 1.417 models from Hugging Face could not find a correlation between model performance and carbon emissions of ML models \cite{JoelESEM23}. This shows the potential for lightweight AI models architectures \cite{castanyer2024design,martinez2023towards}.

Another model-centric strategy is the compression of existing models (aka AI model optimization): structured and unstructured pruning, quantization and binarization, efficient training and inference of models under different resource constraints.
In this regard, Pytorch and TensowFlow are already offering AI models optimization libraries \cite{tf_optimization, pytorch_optimization}. 

\subsection{System-centric Green AI}\label{sec:system}
System-centric green AI refers to the decisions we can take regarding the software architecture and serving of ML systems to make them environmentally sustainable.
The ecological footprint of an AI model extends beyond its algorithmic design to encompass the entire system infrastructure that supports it. For instance, suboptimal hardware choices can significantly inflate energy consumption~\cite{del2023dl}), while implementing batched inference can enhance energy efficiency ~\cite{yarally2023batching}.
It is essential to recognize the continuum from cloud to edge computing in considering energy footprints~\cite{del2023review}. This continuum encompasses end-user devices (edge), cloud servers, and the network infrastructure that connects them. Each component plays a role in determining the overall environmental impact of AI systems, highlighting the need for holistic considerations in system design and deployment for sustainability.

We anticipate that sustainability challenges will initially be tackled with a system-centric approach, given the overlap between AI systems and traditional software systems. However, we argue that emerging challenges lie in the data-centric and model-centric domains of AI, where SE plays a critical role~\cite{shome2022data,shome2024data}. The significance of data quality has already been acknowledged for its impact on the reliability, robustness, efficiency, and trustworthiness of modern software systems~\cite{nguyen2023software}. From a model-centric perspective, there is an unexplored potential to employ SE to challenge the status quo. For example, adopting energy-concious practices for model adaptation~\cite{poenaru2023maintenance}, hyperparameter tuning~\cite{yarally2023uncovering}, and so on.

\section{Reflections on the future of SE}

When reflecting about the future of SE, the emergence of Green AI introduces several challenges that warrant reflection. In this section, we delve into these potential challenges, providing \textit{context} and outlining their \textit{implications for SE}.


{\bf 1. Consideration of the business case.} {\it Context.} Not all problems are alike. Diverse business cases require different positioning with respect to environmental sustainability. An image processing model for cancer diagnose shall prioritize precision no matter what is the carbon footprint required for training. On the contrary, a film affinity recommender system will hardly be considered life-critical, therefore little gains of accuracy at the cost of dramatic increments of energy consumption is not justified; instead, digital sobriety\footnote{\url{https://www.vox.com/climate/2024/3/28/24111721/ai-uses-a-lot-of-energy-experts-expect-it-to-double-in-just-a-few-years} (visited on April 5, 2024)} should start to be the norm, and not the exception.  
{\it Implications for SE.} Business case elaboration and analysis lies at the heart of the requirements engineering area \cite{franch2023requirements}. Its application to AI systems, often denoted by RE4AI, is currently a mainstream in SE, yielding to a number of actionable findings, e.g., definition of new scopes for requirements~\cite{Siebert2020} or new quality requirement types~\cite{Horkoff23}. However, as Habibullah et al. remark, there is a research gap in establishing trade-offs among quality requirements~\cite{Horkoff23}. Understanding and specifying these trade-offs would help to translate the business case into quantifiable requirements, associated metrics, and optimization functions allowing for a proper validation of the AI system.

{\bf 2. Consolidation of fundamental concepts.}
{\it Context.} Design of AI systems is still a young research area. As a consequence, fundamental concepts are still not fully understood and lead to inconsistent or even wrong terminology. For instance, we may find papers using the terms `energy efficiency' with the meaning of `energy consumption', or defining metrics with an incorrect measurement unit. This situation hampers communication and interdisciplinary collaboration~\cite{Lewis23}.
{\it Implications for SE.} Although there are several excellent works that focus on establishing a consolidated terminology for environmental sustainability~\cite{guldner2024development}, the truth is that they have not made their way in the SE community. Observing what happens in other areas, a standard playing the same role as for instance ISO/IEC 25010 in the field of software quality~\cite{ISO25010} could be a significant asset towards this consolidation. Eventually, such a standard could be the upcoming ISO/IEC 20226 on AI environmental sustainability\footnote{\url{https://www.iec.ch/blog/importance-sustainable-ai} (visited on April 5, 2024)}.

\textbf{3. Monitoring sustainability.}
\textit{Context.}
Collecting energy data is not a trivial task. Even after collecting it, practitioners have to process and analyze energy data and tracing it back to the AI pipelines, to hotspots and make necessary adjustments if possible. To make matters worse, improving sustainability is often a trade-off problem: we want to reduce energy consumption without hindering other requirements of the system. For example, improving energy efficiency at the cost of privacy is likely unacceptable, depending on the use case. On another note, the impacts of AI systems on the environment can hardly be simplified by looking at energy consumption alone. For the same energy consumption of executing a AI software, the carbon footprint can vary depending on the time, region, and the type electricity used to power the servers at that particular. Besides, carbon footprint, we also have the problematic of water footprint -- i.e., the water being evaporated due to the software execution; mostly due to cooling down servers. Moreover, embodied carbon footprint also plays an important role. GPU usage needs to be optimized to make sure hardware is not stalling idle most of the time during their lifetime~\cite{wu2021sustainable}.
\textit{Implications to SE.}
There is an open challenge of developing user-friendly energy monitoring tools that seamlessly integrate with various environments, including Edge AI devices, and virtual ones like Docker containers. These tools should not only enable detailed analysis but also offer a comprehensive overview of the energy consumption associated with software products.  Additionally, they need to accommodate diverse resource metrics and provide actionable insights to facilitate environmentally-conscious decision-making throughout the development and maintenance phases of these systems~\cite{guldner2024development}.

{\bf 4. Clarification of roles' involvement.} 
{\it Context.}
The emergence of AI demands knowledge and skills that go beyond those traditionally required in classical SE. The consideration of new roles and the interactions among them have been subject of investigation. Specially, the fit of data scientists in the SE team has been thoroughly investigated~\cite{Zimmermann17,Vogelsang19}. In the context of Green AI, knowledge on environmental sustainability is a must for the optimal consideration of carbon footprint in system design. However, experts on sustainability are hardly involved in the development of AI systems 
{\it Implications for SE.}
Same as sometimes software engineers criticize data scientists for building code not adhering to SE principles, software engineers may be criticized by addressing environmental sustainability without having profound expertise in the team. The role of environmental sustainability expert should be explicitly recognized in the Green AI development team, same as data scientist or domain expert are. This role can be played either by real sustainability professionals, or by software engineers who have acquired along time the necessary knowledge and skills. 

{\bf 5. Changes in the ML lifecycle.}
\textit{Context.}
There has been ongoing efforts to reshape the development lifecycle of modern software systems to cover the inclusion of AI-enabled features. However, along this push to redefine the development lifecycle of AI-enabled software, environmental sustainability is often overlooked~\cite{haakman2021ai,lwakatare2020devops,rzig2022empirical,kreuzberger2023machine}.
\textit{Implications.}
It is quintessential to adding sustainabilty concerns across the whole lifecycle, from the very early stages until the retirement of software and/or respective AI models.
There is an unexplored potential to employ SE to challenge the status quo. For example, model retraining is still the default approach for model adaptation, leaving out other strategies that may be more cost-effective and sustainable~\cite{poenaru2023maintaining,poenaru2023retrain,poenaru2023maintenance}. Moreover, adhering to the three R's principle—reduce, reuse, and recycle (in that specific order)—is essential in fostering sustainability practices within AI systems~\cite{bristy2023green}. We need frameworks that help practitioners 1) opt for competitive alternatives to AI when available (reduce), 2) consume existing models instead of training their own (reuse), 3) monitor model degradation and consider model adaptation strategies (recycle) before disposing a model and training a new one~\cite{poenaru2023maintenance}.

{\bf 6. Quest for open science.} 
{\it Context.} Nowadays, the SE community is fully aware of the importance of open science as a basic principle supporting transparency and replication~\cite{Shull08}. Initiatives at all levels (journals, conferences, national research programs, ...) push the general adoption of open science by researchers. In the AI field, it has become customary to give access to (preprint of) papers, data, software and models in public repositories such as arxiv, Zenodo, GitHub and Hugging Face.
However, there is a lack of clear guidelines on what type of information to store in these repositories. For instance, in the case of Green AI, Castaño et al. report that carbon emission related information is only marginally reported in the most popular model repository, Hugging Face~\cite{Castan2023}. This means that in spite of community willingness to go open, it is still difficult to replicate experiments or to conduct long-term cohort studies.
{\it Implications for SE.}  The research community shall produce consolidated and agreed guidelines to allow software engineers be systematic and rigorous in the documentation of the sustainability dimension of their experiments. Two different actions along this direction are needed: {\it (i)} determine what is the information related to environmental sustainability to be described, {\it (ii)} creating domain-specific description sheets that can be used to consolidate such information. An example in this direction is the ID-card proposed by Abualhaija et al. in the NLP domain~\cite{SallamTOSEM}. Another example focusing on environmental sustainability is GAISSALabel, which gathers a repository of training and inference emissions footprints of ML models \cite{duran2024gaissalabel}. Alternatively, the information can be coded as meta-data in the repositories, although usually this will result in less comprehensive descriptions.


\textbf{7. The role of Education}
\textit{Context.}
Equally important is the imperative to educate the next generation of AI developers in sustainable AI literacy. Green SE is still a niche topic~\cite{cruz2022five}, with very few available materials to support educators. Moreover, the rapidly evolving landscape of AI technologies presents challenges in adapting AI and Computer Science curricula to incorporate Green AI practices. In fact, many Green AI practices are very recent and experimental~\cite{jarvenpaa2023synthesis}.
\textit{Implications for SE.}
The education of SE must explicitly incorporate topics on responsible AI and, more specifically, the environmental considerations of AI systems. Educational materials should be openly accessible to empower aspiring software engineers to prioritize sustainability within their organizations.

{\bf 8. Construction of theories.} 
{\it Context.} There is a large corpus of research works that report experiments of any kind with an environmental sustainability dimension. In fact, the original set of 98 studies considered by Verdecchia et al. in their 2023's paper~\cite{Verdecchia23} is falling short today after only a couple of years. While every individual experiment may be valuable per se, their results still remain mainly de-aggregated and are difficult to be integrated into a holistic body of knowledge, i.e., a theory~\cite{Sjoberg2008}. 
{\it Implications for SE.}  Software engineers shall gradually adopt what Stol and Fitgerald call `a theory-focused research approach'~\cite{Stol13}. Building theories is not only important because of the outcome, but also a theory construction process brings the need of critical thinking to researchers and make them aware of the need of better reporting their experiments as to allow building theories on solid basis. A critical point is the consideration of context~\cite{Dyba12}, determined by the independent and confounding variables of the experiment. Modeling context in the theory may facilitate, for instance, integrating the source of energy in the calculation of carbon emissions.


\section{Green foundation models for SE.}

The usage of large language code models is a powerful tool that is becoming more and more popular. It has been reported recently by Microsoft that, amongst GitHub Copilot users, 40\% of the code they commit is `AI-generated and unmodified'\footnote{Scott Guthrie, Executive of the Cloud and AI group at Microsoft, as of  April 5, 2024: \url{https://www.microsoft.com/en-us/Investor/events/FY-2023/Morgan-Stanley-TMT-Conference}}. Foundation models open the stage for developing software systems without developers creating a single line of code. Whole software systems can soon be generated from natural language prompts.

A wide adoption of foundation models in coding activities poses an important question on the impact it has on the environmental sustainability of SE. On one hand, there is a clear reduction of number of developers that can lead to smaller carbon footprints inherited from having teams working together. On the other hand, there is a concerning energy overhead from continuously prompting code generation models to converge to a version of the software that can go to production. Furthermore, it is not clear whether generated code will abide by energy efficiency coding practices. State of the art code generation models are purely statistical: we argue that the most frequent coding patterns are not necessarily the most energy efficient.

Foundation models ought to provide transparency in terms of sustainability indicators at different stages of the software lifecycle. Such indicators should paint a clear picture on the ecological footprint of training those models, using them, but also using their outputs. Strategies should be adopted to make sure these models are more likely to learn energy-efficient coding practices than regular ones. 

\bibliographystyle{ACM-Reference-Format}
\bibliography{references}


\end{document}